\PassOptionsToPackage{unicode=true}{hyperref} 
\PassOptionsToPackage{hyphens}{url}
\documentclass[]{article}
\usepackage{lmodern}
\usepackage{amssymb,amsmath}
\usepackage{ifxetex,ifluatex}
\usepackage{fixltx2e} 
\ifnum 0\ifxetex 1\fi\ifluatex 1\fi=0 
  \usepackage[T1]{fontenc}
  \usepackage[utf8]{inputenc}
  \usepackage{textcomp} 
\else 
  \usepackage{unicode-math}
  \defaultfontfeatures{Ligatures=TeX,Scale=MatchLowercase}
\fi
\IfFileExists{upquote.sty}{\usepackage{upquote}}{}
\IfFileExists{microtype.sty}{%
\usepackage[]{microtype}
\UseMicrotypeSet[protrusion]{basicmath} 
}{}
\IfFileExists{parskip.sty}{%
\usepackage{parskip}
}{
\setlength{\parindent}{0pt}
\setlength{\parskip}{6pt plus 2pt minus 1pt}
}
\usepackage{hyperref}
\hypersetup{
            pdftitle={Deconvolution of isotopic pattern in 2D-FTICR Mass Spectrometry of peptides and proteins},
            pdfborder={0 0 0},
            breaklinks=true}
\usepackage{graphicx,grffile}
\usepackage{graphicx}
\makeatletter
\def\ScaleWidthIfNeeded{%
 \ifdim\Gin@nat@width>\linewidth
    \linewidth
  \else
    \Gin@nat@width
  \fi
}
\def\ScaleHeightIfNeeded{%
  \ifdim\Gin@nat@height>0.9\textheight
    0.9\textheight
  \else
    \Gin@nat@width
  \fi
}
\makeatother

\setkeys{Gin}{width=\ScaleWidthIfNeeded,height=\ScaleHeightIfNeeded,keepaspectratio}%
\ifx\paragraph\undefined\else
\let\oldparagraph\paragraph
\renewcommand{\paragraph}[1]{\oldparagraph{#1}\mbox{}}
\fi
\ifx\subparagraph\undefined\else
\let\oldsubparagraph\subparagraph
\renewcommand{\subparagraph}[1]{\oldsubparagraph{#1}\mbox{}}
\fi

\makeatletter
\def\fps@figure{htbp}
\makeatother

\usepackage[]{natbib}
\usepackage{amsmath,graphicx, pstricks}
\usepackage{amssymb} 
\usepackage{algorithm} 
\usepackage{algorithmic}
\usepackage{mhchem}
\newcommand\ex{\ensuremath{\mathsf{\textbf{x}}}}

\newcommand\eu{\ensuremath{\mathsf{\textbf{u}}}}
\newcommand\ev{\ensuremath{\mathsf{\textbf{v}}}}
\newcommand\en{\ensuremath{\mathsf{\textbf{n}}}}
\newcommand\ee{\ensuremath{\mathsf{\textbf{e}}}}
\newcommand\ey{\ensuremath{\mathsf{\textbf{y}}}}

\newcommand\eD{\ensuremath{\mathsf{\textbf{D}}}}

\newcommand\Av{\ensuremath{\mathcal{A}}}

\newcommand\ed{\ensuremath{\mathsf{\textbf{d}}}}
\newcommand\cD{\ensuremath{\mathcal{\textit{D}}}}
\newcommand\prox{\ensuremath{\operatorname{prox}}}
\newcommand\Argmin{\ensuremath{\operatorname{Argmin}}}

\newcommand\eR{\ensuremath{\mathbb{R}}}
\newcommand\eN{\ensuremath{\mathbb{N}}}
\newcommand{\minimize}[2]{\ensuremath{\underset{\substack{{#1}}}%
{\mathrm{minimize}}\;\;#2 }}

\title{Deconvolution of isotopic pattern in 2D-FTICR Mass Spectrometry of
peptides and proteins}

\author{
      Laura Duciel(1)
   \and       Afef Cherni(2)
   \and       Marc-André Delsuc(1)
    \footnote{      (1)CASC4DE, Strasbourg, France; IGBMC, Université de Strasbourg, Illkirch,
France
          (2)i2m, CMI, Aix-Marseille University, France
    }
  }

\date{}

\begin{document}
\maketitle
\begin{abstract}
Mass Spectrometry (MS) is a largely used analytical technique in biology
with applications such as the determination of molecule masses or the
elucidation of structural data. Fourier Transform Ion Cyclotron
Resonance MS is one implementation of the technique allowing high
resolution and mass accuracy and based on trapping ions in circular
orbits thanks to powerful Electromagnetic fields. This method has been
extended to two-dimensional analysis, this gives signals containing a
lot more information, for a reasonable amount of time and samples.
However, the data provided by such experiments cannot be stored in a
classical manner and require some tools and architecture to store and
compress data so that they remain accessible and usable without
necessitating too much computer memory. The developed python program
permits to turn these huge raw data into exploitable FTICR MS HDF5
formatted datasets which can be relatively rapidly and efficiently
deconvolved and analysed using a Primal-Dual Splitting algorithm.
\end{abstract}

\hypertarget{introduction}{%
\section*{Introduction}\label{introduction}}
\addcontentsline{toc}{section}{Introduction}

Fourier Transform Ion Cyclotron Resonance Mass Spectrometry (FTICR-MS)
allows to accurately measure the mass over charge ratio (\(m/z\)) of
molecular ions in the gas phase. It is based on a trapping of ions in
cyclotronic circular orbits in vacuum, thanks to a powerful uniform
magnetic field \citep{Marshall:1998ck}. Each of the ions, when excited
by a resonant Radio Frequency electric field will generate a
time-dependent current which can be measured. Fourier Transformation of
this signal allows to get the corresponding orbital frequencies and a
mass spectrum is obtained from a calibration based on the approximate
basic equation:

\begin{equation}
\label{eq:frequency}
f_c = \frac {z B_o}{m}
\end{equation} where \(f_c\) is the cyclotronic resonance frequency,
\(B_o\) is the strength of the static magnetic field, \(m\) and \(z\)
respectively the mass and the charge of the orbiting ion. This technique
affords a high resolution and mass accuracy which eases the molecular
interpretation of detected ions by drastically decreasing the ambiguity
of the possible assignments \citep{Thomas_De_Vijlder_et_al_2017}.

Unambiguous assignments are particularly important when mass
spectrometry is used for a bottom-up characterisation of proteins in a
biological extract. In this approach, all proteins are first partially
enzymatically hydrolysed to peptides, and the peptides from this mixture
are analysed by tandem MS. The mass of each peptides is measured as well
as the mass of its fragments after fragmentation inside the
spectrometer. Due to the complexity of the sample, the classical
approach consists in first separating by high-resolution Liquid
Chromatography, and analysing each peptides sequentially
\citep{Aebersold:2016kt}.

We explore here an alternative route, where the sample is analysed by
two-dimensional FTICR mass spectrometry
(2D-FTICR-MS)\citep{Floris:2018kh}. This global approach which does not
require sample fractionation was proposed long
ago\citep{Pfandler:1987ve} and has known a recent renewal thanks to the
increase in the computer
capacities\citep{Agthoven:2012cx, vanAgthoven2019}. After introduction
of the sample mixture, the fragment ions are collected at the end of a
three pulses sequence that includes the fragmentation delay. The
sequence is built so that the intensity of the fragment ions show a
periodic dependence to the cyclotronic frequency of the parent ion. By
varying a delay in the sequence, this frequency is sampled, and a
bi-dimensional Fourier transform of the data leads to a 2D mass
spectrum, where a fragmentation event leads to a peak located at the
parent \(m/z\) along the vertical (\(F1\)) axis and at the fragment
\(m/z\) along the horizontal (\(F2\)) axis.

\begin{figure}
\centering
\includegraphics{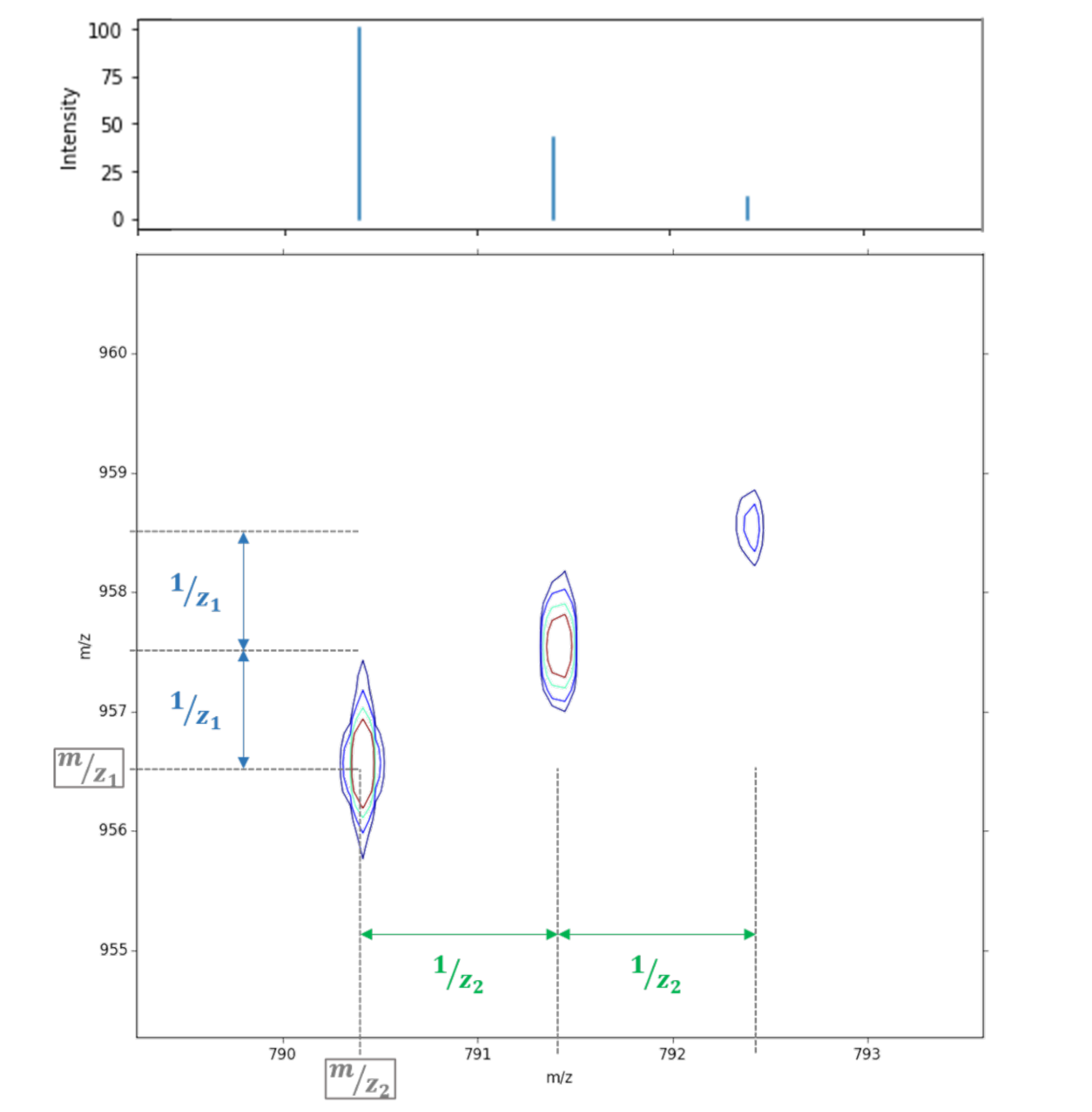}
\caption{(bottom) Zoom on a typical isotopic pattern observed in
2D-FTICR-MS, here from the Yeast digest dataset. Peaks are separated
along each dimensions by the inverse of the charge state \(z_1\) and
\(z_2\), (here \(z_1 = z_2 = 1\)) The monoisotopic peak is located at
the lower left of the diagonal pattern. (top) the theoretical isotopic
pattern computed with the averagine molecular model.
\label{IsotopicPattern}}
\end{figure}

The shape of the signals in a FTICR-MS spectrum is dominated by two
independent phenomena, and is demonstrated in figure
\ref{IsotopicPattern}. The presence of stable isotopes, in particular
\textsuperscript{13}C present at about 1.1\% relatively to
\textsuperscript{12}C, produces typical 2D isotopic patterns, which are
a generalisation of the isotopic patterns classically observed in Mass
Spectrometry, and produced by the statistical distribution of the stable
isotopes in the molecule. The shape of the pattern is easily computed
from the molecular formula and the charge state of the considered ion. A
second phenomena is the experimental peak lineshape, which is dominated
by the experimental details such as the transient duration, and ion
cloud stability, and can be considered as constant for all signals.

A general procedure was recently proposed to perform a pattern
recognition of the classical isotopic pattern
\citep{Cherni-2018-ICASSP}, which efficiently extract the monoisotopic
mass and the charge state. It is based on a dictionary approach. The
goal of this study is to explore the possibility to extend this approach
to the 2D case. It should be noted that most information which has to be
extracted from the 2D experiment is the most possible accurae position
of the monoisotopic signal,

\hypertarget{materials-and-methods}{%
\section*{Materials and Methods}\label{materials-and-methods}}
\addcontentsline{toc}{section}{Materials and Methods}

The raw 2D-FTICR experiment dataset was kindly sent to us by P.O'Connor
from Warwick University. It is a yeast tryptic digest, on which a
regular 2D experiment was performed on a 12 T Bruker Solarix FTICR mass
spectrometer, using ESI ionisation and IRMPD fragmentation. 4096 \(t_1\)
increments were performed, for a total acquisition time of 54 minutes.

The encoding (\(P1\)) and excitation pulses (\(P2\)) were set to go from
\(460\,655\,\)Hz to \(263\,228\,\)Hz, which correspond to a range of 400
to \(700\,m/z\). The measurements along \(F_1\) were made at a Nyquist
frequency of \(fn\) = \(500\,000\,\)Hz. The observed pulse (\(P3\)) was
set to span the frequencies from \(1\,250\,095\,\)Hz to \(61\,418\,\)Hz,
which correspond to a range of 147.4 to \(3000\,m/z\). The acquisition
spectral width was set to \(1\,250\,095\,\)Hz, corresponding to a low
mass limit of \(147.4\,m/z\).

\hypertarget{computer-set-up}{%
\subsection*{Computer set-up}\label{computer-set-up}}
\addcontentsline{toc}{subsection}{Computer set-up}

The analysis and computations were performed on a computer equipped with
a processor Intel Xeon(\textregistered) CPU E3-1240 v5 at \(3.50\,\)GHz
with 4 cores, 16GB of RAM and a graphical unit NVIDIA
Quadro(\textregistered) K1200 with 4GB dedicated RAM.

\hypertarget{data-processing}{%
\subsection*{Data Processing}\label{data-processing}}
\addcontentsline{toc}{subsection}{Data Processing}

The processing of the raw data obtained from the spectrometer were
performed using the SPIKE library after being denoised with the SANE
algorithm. SPIKE is a python library which is available on a public
repository
(http://www.bitbucket.org/delsuc/spike)\citep{chiron_spike_2016}. It
allows to perform the complete processing of a raw 2D-FTICR-MS
measurement, including the import, apodisation, peak-picking methods,
HDF5 format compression. The deconvolution over compressed data is based
on an algorithm designed for 1D deconvolution and extended to 2D, which
is described in the
\protect\hyperlink{theory-and-calculation}{\textbf{Theory and
Calculation}} section.

\hypertarget{experimental}{%
\section*{Experimental}\label{experimental}}
\addcontentsline{toc}{section}{Experimental}

\hypertarget{details-on-the-hdf5-structure-of-the-datasets}{%
\subsection*{Details on the Structure of the
datasets}\label{details-on-the-hdf5-structure-of-the-datasets}}
\addcontentsline{toc}{subsection}{Details on the Structure of the
datasets}

2D FTICR MS datasets are 2D maps, stored as matrices, and produced by a
double Fourier transform step. Each entry is located at \((F_1,F_2)\)
coordinates, expressed in \(m/z\). While being 2D, this is not an image,
and values are stored as double precision floating point values;
visualisation is done by the computation of contour lines.

The 2D FTICR MS dataset is stored into a structured hierarchical HDF5
file. The data is a 2D matrix, and its size prohibits to load the data
in memory, the HDF5 format allows to map the file content to memory, and
this mode was chosen to access the
data\citep{hdf5, dougherty_unifying_2009}. The data matrix has to be
accessed in a random manner in lines, columns, but also as 2D
sub-domains. In a first level of structure, the matrix is stored using
the CARRAY format proposed in HDF5. This format stores the data as
sub-regions, stored as elementary piece of information, which allows an
efficient random access, provided that all the required sub-regions can
fit into memory. Extensive tests have been performed to optimize the
size of these pieces.

However this organisation does not allow to easily produce global views
of the data as it would be extremely slow. For this reason, in a second
level of structure, a set of down-sampled matrices are computed from the
original matrix, and store along side the original, using the
possibilities of the HDF5 format. Several levels of down-sampling is
used, allowing to have rapid and accurate accesses to the whole 2D
spectrum, as well as large or local zooms. In the case of the dataset
analysed here, the original matrix is 4k \(\times\) 256k, and there are
4 down-sampled matrices, with sizes of 1k \(\times\) 64k, 1k \(\times\)
16k, 1k \(\times\) 4k, and 1k \(\times\) 1k, summing to more than a
billion entries and a potential 9.3 Gigabyte of storage. The smaller
matrix allows a fast access to the whole spectrum, while the
intermediate sizes are used for views computed only on a part of the
spectrum. A simple handling program allows to zoom in and out of the
dataset, while always choosing the optimal matrix to use for the
display.

As the dataset is composed of sharp peaks, located in an otherwise empty
background, an optional further improvement has been implemented. A
smaller file can be obtained by setting all values below a certain
threshold to zero, and activating the internal lossless compression
available in the HDF5 format. The threshold is typically chosen after an
estimate of the background noise \(\sigma\). In the case of the dataset
analysed here, a threshold of \(1 \times \sigma\) produced a modest 17\%
compression without no visible modification of the spectrum, and a
\(3 \times \sigma\) threshold gave a 82\% compression, with some of the
smallest features missing. With this set-up, dataset up to 8k x 512k are
routinely handled on a simple laptop computer.

\hypertarget{core-code}{%
\subsection*{Core Code}\label{core-code}}
\addcontentsline{toc}{subsection}{Core Code}

The program development is done in python, a programming language
conceived to be powerful but also easy to learn and to apply. The
program relies on the following libraries \texttt{NumPy} for array
processing\citep{Walt2011}, \texttt{SciPy} for scientific computing
\citep[\citet{bressert_scipy_2012}]{Jones2001}, \texttt{pytable} for
HDF5 files handling\citep{Tables}. We use \texttt{SPIKE}
\citep{chiron_spike_2016} a library developed for the analysis of NMR
and MS data, for the general framework of the program.

\hypertarget{implementation-on-the-whole-2d}{%
\subsection*{Implementation on the whole
2D}\label{implementation-on-the-whole-2d}}
\addcontentsline{toc}{subsection}{Implementation on the whole 2D}

To implement the deconvolution algorithm on the whole 2D dataset at the
highest possible resolution, it is necessary to perform the calculations
by chunks due to the huge size of data. To get these chunks, we use the
memory mapping mode allowed by the \texttt{pytable} library, which avoid
the loading of the complete dataset.

From this zone of interest to analyse,~a list of the \(N\) chunks
coordinates of fixed sized in points is generated. In our set-up, chunks
of \(3072 \times 256\) points were the most efficient as the processing
is relatively fast and the size is large enough to accommodate full
isotopic patterns (see Table \ref{tab:deconv_times}).

For each zone, the deconvolution pattern is computed with a sliding
approximation as presented below, using the averagine model
\citep{senko1994} to estimate a typical molecular formula compatible
with the local \(m/z\), and a fast algorithm to estimate an isotopic
pattern \citep{yergey1983general, Kubinyi:1991}.

The deconvolution is then performed on each chunk on a slightly larger
zone (\(\pm 2 m/z\)~unit) than the chunk coordinates, in order to avoid
any loss of information from patterns potentially cropped by the chunk
border. At the end of the processing, the results is stored in a HDF5
file and memory is freed. The deconvolution results can then be loaded
altogether to get a completely deconvolved spectra. The complete
implementation of the program can be found at
\href{http://www.github.com/LauraDuciel/MSDeconv}{www.github.com/LauraDuciel/MSDeconv}

\hypertarget{theory-and-calculation}{%
\section*{Theory and Calculation}\label{theory-and-calculation}}
\addcontentsline{toc}{section}{Theory and Calculation}

\hypertarget{Pb}{%
\subsection*{Problem formulation}\label{Pb}}
\addcontentsline{toc}{subsection}{Problem formulation}

Let us consider a MS spectrum, presented as a sequence of
\((\{m_p^{\text{iso}}/z_p, a_p\})_{1 \leq p \leq N}\) where we denote by
\(m_p^{\text{iso}} \in (0,+\infty)\) the monoisotopic mass,
\(z_p \in \eN^*\) design the charge state and \(a_p \in (0,+\infty)\) is
the abundance of the peptide \(p \in \left\{1,\ldots,P\right\}\) in the
chemical sample. We propose to model the acquired MS spectrum \(y\) as
the weighted sum of each individual isotopic pattern
\(y = \sum_{p=1}^P a_p \cD(m_p^{\text{iso}},z_p) + n\) where \(n\)
models the acquisition noise and possible errors arising from the
spectral analysis preprocessing step. The measurements are taken on a
discrete grid of \((m/z)\) values with size \(M\), so that the
observation model finally reads:

\begin{equation}
\label{eq:model2}
\ey = \sum_{p=1}^P a_p \ed(m_p^{\text{iso}},z_p) + \en
\end{equation} with \(\ey \in \eR^M\),
\(\ed(m_p^{\text{iso}},z_p) \in [0,+\infty[^M\) and \(\en \in \eR^M\).
The set of coefficients
\((a_p, m_p^{\text{iso}}, z_p)_{1 \leq p \leq P}\) is not easily at hand
because of the complicated and nonlinear function \(\ed(m,z)\) and the
large value of \(M\). The averagine model proposed by \cite{senko1994}
can be defined as a \(\Av : \, \eR \rightarrow \eN^5\) mapping between a
mass value and the number of each atom present in the protein formula
\(\ce{C}_{N_{\ce{C}}}\ce{H}_{N_{\ce{H}}}\ce{O}_{N_{\ce{O}}}\ce{N}_{N_{\ce{N}}}\ce{S}_{N_{\ce{S}}}\).
The result of averagine model on a given mass is a set of positions and
their correspondent intensities presenting the abundances of present
isotopes with their positions along the mass axis. We propose to use
this model to build a dictionary-based approach under the assumption
that we know approximately the range of mass and charge state for the
\(P\) proteins present in the sample. In general, the theoretical
isotope distribution follows a multinomial distribution as it is
presented in \cite{kaur2004use} and can be represented with Gaussian or
Lorentzian shapes. Consequently, the mass distribution function
\(\cD(m,z)\) is easy to evaluate from the molecular formula at a given
\((m,z)\) values. In our work, we propose to generate \(\cD(m,z)\) in
the time domain with a Gaussian shape in order to have a sampled version
\(\ed(m,z)\) on a defined mass grid. Then we use Fourier transform to
map the spectrum domain. Here, we propose to normalize the result, so as
to preserve the sum of squared amplitudes from \(\cD(m,z)\) to
\(\ed(m,z)\).

\hypertarget{isotopic_deconvolution}{%
\subsection*{Isotopic deconvolution}\label{isotopic_deconvolution}}
\addcontentsline{toc}{subsection}{Isotopic deconvolution}

For a given mass axis \([m_{\text{min}},m_{\text{max}}]\), and different
charge states \(z_i \in [z_{\text{min}},z_{\text{max}}]\), we propose to
define a search grid with size \(T:= M Z\) which defines \(M\) possible
values of isotopic masses and \(Z\) possible values for the charges.
From this grid, we build the dictionary \(\eD \in \eR^{M \times T}\)
noted as
\(\eD = \left[\eD_1 \left|\ldots\right|\eD_{\ell}\left|\ldots\right| \eD_Z\right]\)
where, for every \(\ell \in \left\{1,\ldots,Z\right\}\), the sub-matrix
\(\eD_{\ell} \in \eR^{M \times M}\) maps for the dictionary associated
to charge \(\ell\). And each \(i-\)th column of \(\eD_{\ell}\) is
considered as the isotopic pattern distribution \(\ed(m_i,z_i)\)
depending on the mass \(m_i\) and charge \(z_i\) for every
\(m_i \in [m_{\min},m_{\max}]\) and \(z_i \in [z_{\min}, z_{\max}]\).
Then, the problem is reformulated as:

\begin{equation}
\label{eq:dictionnary_model}
\ey = \eD \ex + \en'
\end{equation} where \(\ex\) is a sparse vector with positive entries,
for which the \(P\) non-zeros coefficients allow to determine the
isotopic mass and charge state of each protein, along with their
abundance. Moreover, \(\en' = \en + \ee\) models the acquisition noise
and possible errors arising from the spectral analysis and
discretization steps (\(\en' \rightarrow \en\) with high accuracy). With
the ill-conditioning properties of dictionary \(\eD\) and the presence
of noise, our problem is an inverse ill-posed problem. In addition, the
large size of experimental MS spectrum requires efficient data
processing algorithms, able to handle efficiently the large data sets
involved. Based on the model in (\ref{eq:dictionnary_model}), the
computation of \(\eD\) presents a challenge as large memory resources
are needed to store this matrix. Therefore, we propose a new
approximation based on Fourier transform. For similar mass values,
isotopic patterns differ merely by a simple translation of peaks
positions, and isotopic patterns can thus be considered locally stable,
and the problem can locally be expressed as a strict convolution To
avoid the storage of isotopic patterns for each \(m/z\) value required
by the dictionary-based approach, we decompose the mass axis into small
windows onto which the isotopic pattern is assumed to be constant up to
a circular shift. Let \(L \leq M\) the chosen window width and
\(\overline{\ed}_{s,\ell}\) the average isotopic pattern for a mass
within the range \([(s-1) L + 1, s L]\), and a fixed charge state
\(z_{\min} + (\ell - 1) z_{\max}\). We propose to approximate each
\(\eD_{\ell}\) by the following block diagonal (BDiag) matrix made of
\(M/L\) blocks assumed to be circulant (Circ) matrices with first line
\(\overline{\ed}_{s,\ell}\), \(s \in \left\{1, \ldots, M/L\right\}\):

\begin{equation}
\overline{\eD}_{\ell} = \operatorname{BDiag}\left(\left[\text{Circ}\left(\overline{\ed}_{s,\ell} \right)\right]_{1 \leq s \leq M/L} \right).
\label{eq:Dicapprox}
\end{equation}

As a consequence, the circulant approximation will be noted as
\(\overline{\eD} = \left[\overline{\eD}_1 \left|\ldots\right|\overline{\eD}_{\ell}\left|\ldots\right| \overline{\eD}_Z\right]\)
and for every charge value, the products \(\overline{\eD}_{\ell}\) and
\(\overline{\eD}_{\ell}^\top\) with vectors can be easily computed using
Fourier operations.

\hypertarget{Optimization_strategy}{%
\subsection*{Implementation}\label{Optimization_strategy}}
\addcontentsline{toc}{subsection}{Implementation}

A direct inversion of \(\eD\) to solve problem
(\ref{eq:dictionnary_model}) and find an estimate of \(\overline{\ex}\)
is not feasible because of the ill-conditioning character of \(\eD\) and
the presence of noise. Therefore, we propose to employ a penalisation
approach that defines the estimate \(\widehat{\ex} \in \eR^T\) as a
solution of the constrained minimization problem:

\begin{equation}
\label{eq:reg_dic}
\minimize{\ex \in \eR^T}{\Phi(\ex)} \quad \text{subject to} \quad  \| \eD \ex - \ey \| \le \tau
\end{equation} where \(\Phi: \eR^T \mapsto (-\infty,+\infty]\) is the
regularization function used to enforce positivity and sparsity on the
solution, and \(\tau>0\) is a parameter based on an estimate of
experimental noise. The resolution of this problem requires to compute
the proximal operator of function \(\Phi\) at \(x \in \eR^N\), defined
as the unique minimizer of \(\Phi + \dfrac{1}{2} \|\cdot - x \|^2\)
\cite{moreau1965proximite, bauschke2011convex}. This operator has been
generalized for lower semicontinious and proper functions that are not
necessarily convex in \cite[Sec.XV-4]{hiriart1993convex}, as the
multi-valued operator:

\begin{equation}
\prox_{\alpha \Phi} : \ex \mapsto \underset{\ey \in \eR^N} {\Argmin} \quad \left(\frac{1}{2\alpha}\|\ey-\ex\|^2 + \Phi(\ey)\right).
\label{eq:ProxDef}
\end{equation}

Using the variational formulation (\ref{eq:reg_dic}) we propose to use
the proximal Primal-Dual Splitting algorithm from
\cite{chambolle2011first} which is an efficient algorithm for convex
optimization and we choose to use the \(\ell_1\) norm as a sparsity
penality.

\begin{algorithm}[H]
\caption{Primal-Dual Splitting Algorithm}
\label{alg:PD}
$$
\left\lfloor
\begin{array}{l}
\text{\bf Initialization}\\
\eu^{(0)} \in \eR^{M}, \, \ex^{(0)} \in  \eR^T \\
0 < \nu \kappa \|\eD\|^2 < 1, \, \kappa>0 \\
\rho \in (0,2)\\
\text{\bf Minimization}\\
\text{For}\;\;k = 0,1,\ldots\\
\left\lfloor
\begin{array}{l}
\tilde{\ex}^{(k)} = \prox_{\kappa \Phi}(\ex^{(k-1)} - \kappa\eD^{\top}(\eu^{(k-1)}))\\
\ev^{(k)}         = \eu^{(k-1)} + \nu \eD(2\tilde{\ex}^{(k)} - \ex^{(k-1)})\\
\tilde{\eu}^{(k)} = \ev^{(k)} - \nu \text{proj}_{\| \cdot - \ey \| \leq \tau}(\ev^{(k)} / \nu)\\
\ex^{(k)} = \ex^{(k-1)} + \rho (\tilde{\ex}^{(k)} - \ex^{(k-1)})\\
\eu^{(k)} = \eu^{(k-1)} + \rho (\tilde{\eu}^{(k)} - \eu^{(k-1)})\\
\end{array}
\right.
\end{array} 
\right.
$$
\end{algorithm}

Here above, the projection operator
\(\text{proj}_{\| \cdot - \ey \| \leq \tau}\) is defined, for every
\((\ey,\ev) \in (\eR^N)^2\), as:

\begin{equation}
\text{proj}_{\| \cdot - \ey \| \leq \tau}(\ev) = \ev + (\ev - \ey) \min \left(\frac{\tau}{\|\ev - \ey\|},1 \right) - \ey.
\end{equation}

The convergence of the iterates \((\ex^{(k)})_{k \in \eN}\) to a
solution of problem \eqref{eq:reg_dic} is ensured, according to
\cite{chambolle2011first} and \cite{Condat2013}. To estimate the mass
and charge positions, Primal Dual algorithm can be easily used with the
dictionary-based approach. Moreover, the 2D MS spectra can be
efficiently analysed with a sample matricisation step. In this case, the
convolution approximation approach can been used with little
modifications where \(\eD\) is replaced by \(\overline{\eD}\)
\eqref{eq:Dicapprox} and norm of \(\overline{\eD}\) is computed using
power iteration.

\hypertarget{optimization}{%
\subsection*{Instrument response}\label{optimization}}
\addcontentsline{toc}{subsection}{Instrument response}

In this process the physical model \(\ed(m,z)\) and its implementation
\(\overline{\eD}\), comprehend the theoretical isotopic pattern and the
instrument response function, which determines the width of the measured
signal. Adding the instrument response to the model allows a
deconvolution of the signal, and produces an enhancement of the
linewidth along with the analysis of the \(z\) value. Determination and
optimization of the instrument function is required.

\begin{figure*}[!htp]
\centering
\includegraphics{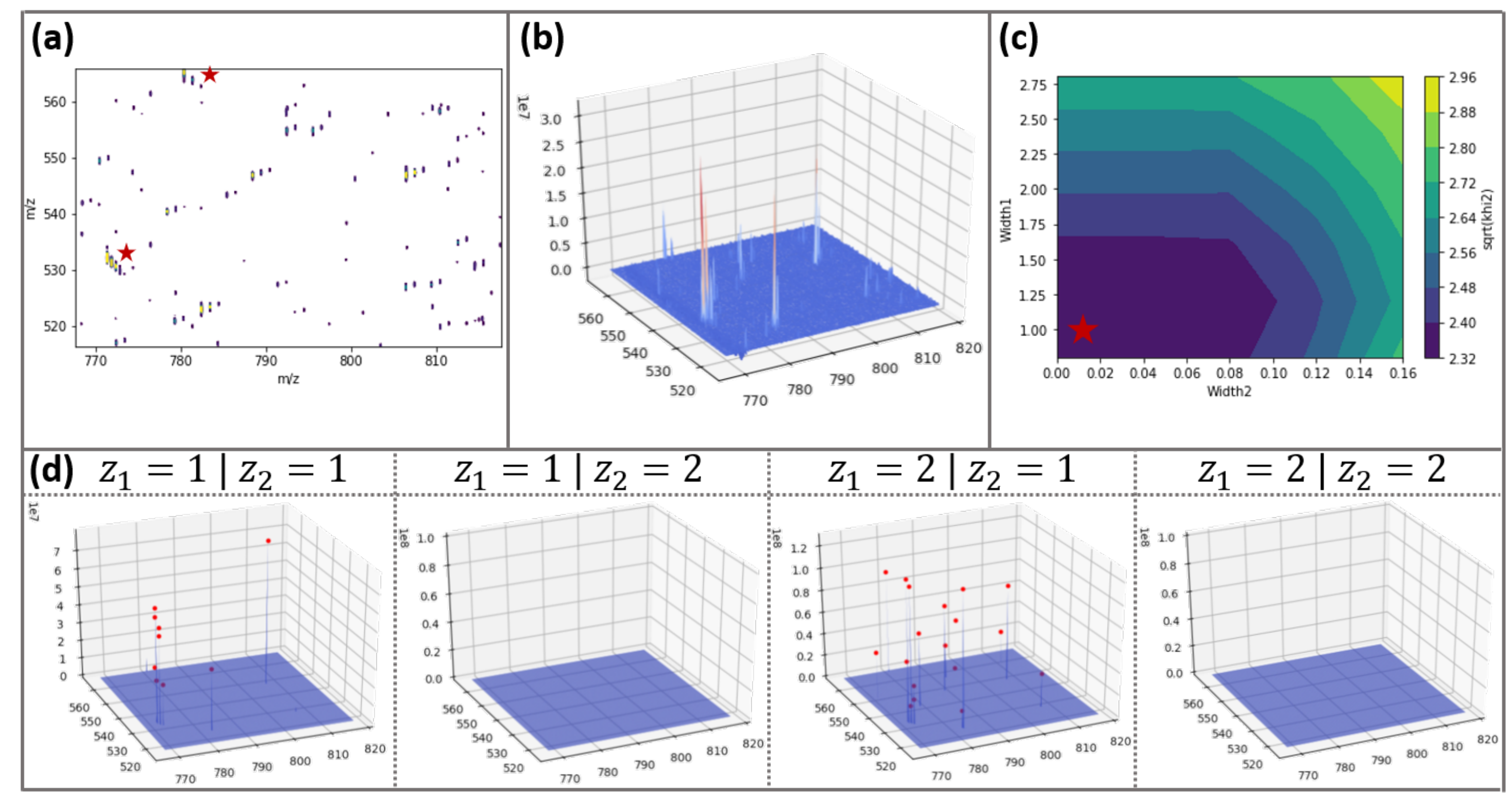}
\caption{The deconvolution process,
a) a zoomed region of the whole analysed dataset, the stars indicates signals from folded harmonics (see \cite{vanAgthoven2019});
b) the same region as in a) shown in 3D; 
c) the value of $\sqrt{\chi^2}$ obtained when varying the width parameters along both axis, the star indicates the value used for the whole analysis;
d) the result of the optimized deconvolution on the same region for the various $(z_1,z_2)$ pairs.
\label{widthoptimdeconv}}
\end{figure*}

As an approximation, we assume that the instrument produces Gaussian
lineshapes along both dimensions, and this is coded in the definition of
\(\overline{\eD}\). We observed that the linewidth is more or less
constant in frequency. Because of the inverse law between frequency and
\(m/z\) (Eq \ref{eq:frequency}), this implies that the linewidth
\(\Delta m\) is proportional to \((m/z)^2\). Expressed in terms of
resolution \(R = m/\Delta m\), this means that \(R\) is proportional to
\(\frac 1 {m/z}\) so that \(Rm/z\) is a constant over the whole
experiment, it could thus be estimated for some specific zones and
applied globally. The optimal \(Rm/z\) was estimated by varying the
widthes and comparing the value of thr normalized \(\chi^2\) at
convergence, as shown on Figure \ref{widthoptimdeconv}. An optimum close
but larger than \(1.0\) indicates that the Gaussian approximation is
correct but does not fully describes the instrument response. For the
yeast data, it was estimated that \(Rm/z = 2.8\,10^5\) equivalent to
\(\Delta m/m^2 = 3.4\, 10^{-6}\) on the vertical axis and
\(Rm/z = 1.3\, 10^7\) equivalent to \(\Delta m/m^2 = 10^{-7}\) on the
horizontal axis.

\begin{figure}
\centering
\includegraphics{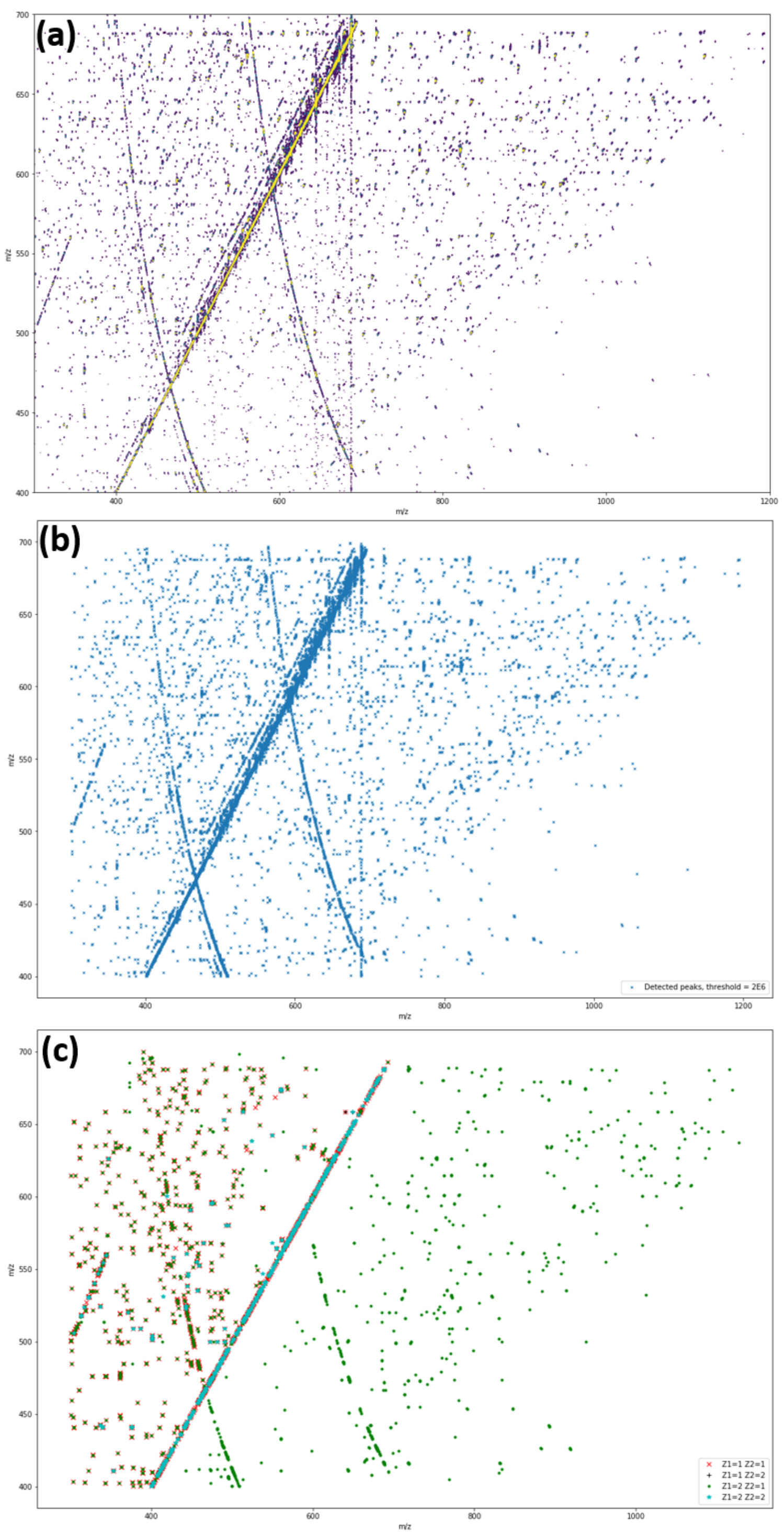}
\caption{The result of the analysis of the complete 2D FTICR experiment
using the optimised procedure. a) the original dataset; b) the result of
the pattern analysis, with the superposition of the 4 \((z_1, z_2)\)
analysed possibilities; c) the automatically detected peak from the
dataset in b), with the superposition \emph{red}:
\((z_1, z_2) = (1,1)\), \emph{black}: \((z_1, z_2) = (1,2)\),
\emph{green}: \((z_1, z_2) = (2,1)\), \emph{blue}:
\((z_1, z_2) = (2,2)\) showing the 4300 detected patterns. Note that
because of physical reasons, only green peaks should be present below
the diagonal.\label{ppbeforeafterdeconv}}
\end{figure}

A peak picking and centroid centering was finally performed on the
resulting reconstructed spectra, and the monoisotopic peak list listed.
Here, 4300 peaks were retrieved after the deconvolution process and
cleaning of impossible cases of \((z_1,z_2)\) pairs, to be compared with
the 11078 peaks observed on the non-deconvoluted dataset. (Figure
\ref{ppbeforeafterdeconv}) This peak list can then directly be used for
database search for the recognition of proteins such as Mascot or
Prosight.

\hypertarget{results}{%
\section*{Results}\label{results}}
\addcontentsline{toc}{section}{Results}

The application of this procedure on a yeast extract sample is presented
in Figures \ref{widthoptimdeconv} and \ref{ppbeforeafterdeconv} and in
Table \ref{tab:deconv_times}. Pattern recognition and monoisotopic
analysis is successfully performed in a reasonable time considering the
size and complexity of the dataset.

\hypertarget{computing-times}{%
\subsection*{Computing times}\label{computing-times}}
\addcontentsline{toc}{subsection}{Computing times}

The deconvolution of a 2D chunk is very dependent on the size of the
chunks. Speed tests have been realized and show that the algorithm is
more efficient on smaller chunks (Table \ref{tab:deconv_times}).
However, care has to be taken to insure chunks large enough to
accommodate a complete pattern.

\begin{table*}[]
\centering
\begin{tabular}{|c|c|c|c|}
\hline
\multicolumn{1}{|p{3cm}|}{\centering Chunk Sizes} 
& \multicolumn{1}{|p{3cm}|}{\centering Processing time observed for 1 chunk}
& \multicolumn{1}{|p{3cm}|}{\centering Processing time expected for the whole experiment} 
& \multicolumn{1}{|p{3cm}|}{\centering Processing time observed for the whole experiment} \\ \hline
$1024 \times 256$          & 1 min 15 sec         & 18.3 h        & 12.6 h \\ \hline
$1536 \times 256$      & 2 min 30 sec         & 24.5 h        & 14.8 h \\ \hline
$2048 \times 256$        & 4 min 30 sec         & 33 h          & 15 h \\ \hline
$3072 \times 256$        & 10 min 30 sec        & 51.5 h        & 20.4 h \\ \hline
$2048 \times 512$      & 16 min 30 sec        & 70 h          & not evaluated \\ \hline
$3072 \times 384$    & 20 min               & 70 h          & not evaluated \\ \hline
\end{tabular}
\caption{Deconvolution processing times for various chunk sizes;
observed processing time on the whole experiment include the $2m/z$ safety border}
\label{tab:deconv_times}
\end{table*}

Taking into account these parameters, we estimated that the best chunk
size choice would be \(3072 \times 256\), taking approximately
\(10\,\)min \(30\,\)sec for the deconvolution of one chunk for
\(z \in [1,2]\) on the displayed zone. This selected chunk size offers
good performances in terms of time and deconvolution efficiency. It was
also observed that the deconvolution on highly compressed dataset was
presenting artifacts and was clearly less efficient, probably due to bad
convergence of the algorithm on zones containing too much zeros after
the compression process.

The parameters should also be adapted to the computer set-up on which
the calculations are performed. To get a result over the whole
\(4k \times 256k\) real points (representing more than a billion data
points) with \(3072 \times 256\) chunks it took \(20.4\,\)hours on the
tested computer described in the
\protect\hyperlink{materials-and-methods}{\textbf{Materials and
Methods}} part (Figure \ref{ppbeforeafterdeconv}). It should finally be
noted that this algorithm, decomposing the dataset in piecewise
independent domain, can easily be parallelized on a larger system.

\hypertarget{discussion}{%
\section*{Discussion}\label{discussion}}
\addcontentsline{toc}{section}{Discussion}

Two-dimensional spectrometry such as 2D-FTICR-MS presents an exceptional
analytical power, but introduces new challenges when compared to regular
analytical measurements. Known in 2D Nuclear Magnetic Resonance (2D-NMR)
spectroscopy for more than 30 years, problems such the size of datasets,
the presence of non-local structured patterns, complex lineshapes have
already been tackled and solved. Based on equivalent principles than 2D
NMR, 2D FTICR-MS full development is only recent \citep{Agthoven:2012cx}
and these problems are not only present, but probably more stringent
because of the size of the datasets.

The most important parameter in mass spectrometry is the determination
of the exact position of spectral peaks, which determine the exact mass
of the studied compound, in addition, the theoretical signal is always a
\(\delta\) function with no intrinsic width. For this reason, the
decomposition of the instrument lineshape, as well as the non-local
isotopic pattern to a single sharp peak is of importance. In addition,
the size of the dataset enforces the use of complex data-storage, with
hierarchical down-sampled representation and arrayed storage.

We show here that the methods developed to analyse classical FTMS
spectra can be extended to the 2D case. The use of an optimized
algorithm, based on a simplified dictionary allows to efficiently
express the varying patterns to be recognized. In addition, the use of a
HDF5 physical representation of the dataset insure an efficient data
access, even with a limited memory, while allowing internally compressed
files. The spectrum resulting from this analysis, was stored in the same
manner, and presented very high compression ratio, thanks to the
sparsity enforcing algorithm which generate dataset with many null
values.

With this organisation, we were able to process in a compatible time a
crowded, real life experiment, on a small desktop computer running with
a limited memory and on only one processor. A parallel implementation of
this algorithm was not performed but would be straightforward, and would
insure high quality results in less than an hour when run on a larger
computer or a small departmental cluster.

\hypertarget{conclusions}{%
\section*{Conclusions}\label{conclusions}}
\addcontentsline{toc}{section}{Conclusions}

We have shown here that the method previously proposed for the analysis
of 1D MS spectra and accurate determination of monoisotopic values by
isotopic pattern matching \citep{Cherni-2018-ICASSP}, can be extended to
the 2D FTICR MS experiment. It was implemented and successfully applied
on a real dataset. The code used in this work is open-source, and
available at the following address:
\href{http://www.github.com/LauraDuciel/MSDeconv}{www.github.com/LauraDuciel/MSDeconv}.

\hypertarget{acknowledgements}{%
\section{Acknowledgements}\label{acknowledgements}}

This work has received funding from the European Union's Horizon 2020
research and Innovation programme under grant agreement EU-FTICR-MS No
731077. We thank Peter O'Connor for the gift of the dataset and
enlighting discussion on the project, and Émilie Chouzenoux for advices
and help on the algorithmic development.

\bibliography{bibliography.bib}

\end{document}